# Aggregate-networks and *p*-cores of monoflow partially overlapped multilayer systems

**Olexandr Polishchuk**

Laboratory of Modeling and Optimization of Complex Systems, Pidstryhach Institute for Applied Problems of Mechanics and Mathematics, National Academy of Sciences of Ukraine, Lviv, Ukraine

E-mail: od_polishchuk@ukr.net

**Abstract**
The paper introduces a number of structural and functional features of classification of multilayer networks (MLN), by means of which distinguish monoflow partially overlapped MLN, that are quite common in the study of intersystem interactions of different nature. The concept of MLN's aggregate-network is defined, which in many cases significantly simplifies the study of intersystem interactions, and the properties of its *k*-cores are investigated. The notion of *p*-cores is introduced, with help of which the components of MLN that are directly involved in the implementation of intersystem interactions are distinguished. Methods of reducing the complexity of multilayer network models are investigated, which allow us to significantly decrease their dimensionality and better understand the processes that take place in complex network systems and intersystem interactions of different types. Using the introduced local and global structural characteristics of elements, effective scenarios of targeted attacks on monoflow partially overlapped multilayer networks have been developed, the main attention of which is focused on the transition points of MLN through which intersystem interactions are actually realized. It is shown that these scenarios can also be used to solve the inverse problem, namely, which elements of MLN should be blocked in the first place to prevent the acceleration of spread of epidemics of dangerous infectious diseases, etc.

______________________________________________________________________________

## 1. Introduction

Any real system is open, ie it interacts with other systems [1-3]. Intersystem interactions of different types give rise to different types of interconnected network structures, which in the theory of complex networks (TCN) are called multilayer networks (MLN) [4, 5]. At present, a number of the most common types of MLN in the physical world, wildlife and human society have been identified. First of all these are multidimensional networks, each layer of which reflects a different type of interaction [6, 7]. The sets of nodes of separate layers of multidimensional network usually intersect at least partially. An example of multidimensional MLN is international cooperation, which includes the layers of political, economic, military, security, cultural, sports and other types of interactions between the nodes-countries of the Earth. Each city of the country is a node of several transport networks of different types, state and local government networks, economic and financial networks, etc. The city's life support system includes electricity, gas and water supply networks, telephone networks, cable television, Internet services, etc. Every person is also a node of many networks (family, professional, social, religious, etc.). Initially, the structures of intersystem interactions of different types were displayed in the form of weighted multigraphs, in which each weighted (numbered by type of interaction) edge reflected the corresponding type of interaction [8]. Obviously, this method is quite inconvenient for displaying structures that have a huge number of nodes and many different types of interactions. Multidimensional networks are much more convenient for better visual perception and analysis of the properties of such structures. The next type of MLN are interacting networks, in which each layer reflects the structure of certain type of interaction and the sets of nodes of separate layers generally do not intersect [9, 10]. Interactive networks often arise when several network systems of different types and purposes are forced to cooperate to solve problems important for joint vital activity. For example, overcoming the consequences of natural or man-made disasters, pandemics, terrorist acts, requires the cooperation of rescue and fire services, police and security agencies, army and medical units, etc. Conducting interdisciplinary research requires the cooperation of scientists of various scientific specialties [11]. The human body consists of many different systems (cardiovascular, lymphatic, nervous, musculoskeletal, respiratory, digestive, etc.). Coordinated joint functioning and interaction of these systems actually forms a living organism. Multiplexes are MLN, all layers of which contain the same set of nodes and have a different structure of edges [12, 13]. Partially overlapped multiplex networks [14] are generalizations of multiplexes in which the sets of nodes of different layers intersect only partially. Such MLN more adequately reflect the structure of many really existing intersystem interactions: transport, social, linguistic, and so on [15, 16]. The advantage of partially overlapped multiplex structures is the possibility of access to nodes of one layer, which are inaccessible in another (the use of air transport where there are no roads or railways, water - in regions where there are no airports or helicopters, etc.). Temporal networks [17] are MLN, in which each subsequent layer sequentially describes the structure of the same network system in the process of its development at certain intervals [18]. Multilevel networks [19] are a set of networks, some of which are consistently part of others, ie hierarchically-network structures with a hierarchy of nesting [20]. Interdependent networks [21, 22] are the sets of networks, the nodes of some of which are control for the nodes of others, ie hierarchically-network structures with a hierarchy of direct subordination [20]. Other structures of intersystem interactions are also studied [23, 24]. The above-mentioned MLN, as well as ordinary complex networks (CN), can be binary and weighted, oriented and unoriented, homo- or heterogeneous in terms of nodes and edges [25], etc. As a strict classification of MLN has not yet been developed, different types of structures are often understood under the same notion, and different notions can define similar structures. Thus, multilayer networks are sometimes called networks of layers [4], multidimensional networks - multirelational [26], interacting MLN - interconnected networks [27], temporal – evolving MLN [28], interdependent - layered networks [29], both multilevel and interdependent MLN are often called network of networks [30], although this term is often used to refer to other types of MLN [4]. In the above list of MLN in some cases, one structure may be a partial case of others, for example, each network system in the process of its operation generates a temporal network. Temporal networks, depending on the length of time between the separate layers, may have the structure of a multiplex, partially overlapped multiplex network or interacting MLN. There are intersystem interactions whose structure simultaneously has the properties of several described above MLN (territorial and administrative division in the country, as a combination of multilevel and interdependent networks, etc.). In many publications [31, 32] the type of structure is not defined at all and the general term "multilayer(ed) network" is used. In addition, there are intersystem interactions, which are currently relatively little studied using network approaches, for example structures generated by coupled fields [33, 34], or intersystem influences of different types. Often these influences are "slow" compared to the cascading phenomena in the network, but no less harmful or even destructive. There are supersystem formations that operate under different laws and have a similar structure, however, exist intersystem interactions that operate under similar laws and generate different types of structures (political and administrative governance in democracies, etc.). It should be recognized that at present a clear generally accepted classification of intersystem interactions, at least at the level of MLN, has not been developed. Such situation requires the definition of strict features that would help to create such classification. It turns out that purely structural features are not enough, because it is the system in the process of formation and functioning builds its structure, and not vice versa. It should also be noted that models of intersystem interactions structures often have a huge dimensionality and complexity, associated with both a large number of MLN nodes and amount of various intra- and interlayer interactions between them. This requires the development of methods to reduce the complexity of models of such formations, aimed at simplifying their understanding and cognition [35], as well as overcoming the problem of complexity of systems research in general [36, 37].

## 2. Structural model of multilayer network

Usually, multilayer network structures are described as

$$G^M = \left( \bigcup_{m=1}^{M} G_m, \bigcup_{\substack{m,k=1 \\ m \neq k}}^{M} E_{mk} \right), \qquad (1)$$

where $G_m = (V_m, E_m)$ is a description of structure of the $m$-th network layer of MLN; $V_m$ is a set of nodes of the network $G_m$; $E_m$ is a set of edges of the network $G_m$; $N_m$ is a number of elements of $V_m$; $L_m$ is a number of elements of $E_m$ (taking into account the direction of network edges); $E_{mk}$ is a set of edges between the nodes of sets $V_m$ and $V_k$, $m \neq k$. Here and hereafter we assume that $m, k = \overline{1, M}$, where $M$ is a number of layers of MLN. The set

$$V^M = \bigcup_{m=1}^{M} V_m$$

will be called the total set of MLN nodes, $N^M$ is a number of nodes of $V^M$.

We represent the mathematical model of MLN in the form of adjacency matrix

$$\mathbf{A}^M = \{\mathbf{A}^{km}\}_{k,m=1}^{M}.$$

Blocks $\mathbf{A}^{km} = \{a_{ij}^{km}\}_{i,j=1}^{N^M}$ of this matrix are defined for the total set of MLN nodes, ie the problem of coordination of nodes numbers in the case of their independent numbering for each layer disappears. We consider MLN as a structure that should provide the movement of both unidirectional and bidirectional flows in the process of intersystem interactions, and the volumes of flows between two nodes in different directions can differ significantly. That is, matrix $\mathbf{A}^M$ is generally asymmetric. The location (numbering) of layers in the multilayer structure and the direction of edges between the nodes of these layers are clearly ordered. This ordering greatly simplifies the construction and subsequent use of adjacency matrices for many MLN. In particular, for temporal networks or MLN with sequential (layer-to-layer) connections, the adjacency matrix will have an upper two-diagonal block structure, in which each block determines the interactions between the nodes of respective network layers, for interdependent MLN with a linear control model this matrix will have a three-diagonal block structure etc.

The structure (1) and adjacency matrix $\mathbf{A}^M$ describe the MLN of most general form. Let's define a number of structural and functional features that simplify the classification of multilayer network systems (MNS).

### 2.1. Structural features of multilayer networks classification

On the basis of intersection of the sets of layers nodes, we distinguish the following types of MNS structures.
1. Multilayer structures in which the sets of nodes of all network layers are identical, ie
$$V_m \bigcap V_k = V_m = V_k = V^K.$$
Such structures are called multiplexes [12]. Examples of multiplexes are the above-mentioned networks of international cooperation, intersystem interaction between the settlements and citizens of the country and many others.

2. Partially overlapped (PO) multilayer networks, ie structures in which the sets of nodes of all network layers intersect partially. In this type of MLN we will allocate two subtypes:

   2.1. Partially overlapped multilayer networks in which there is a non-empty intersection of all network layers, ie
$$V_m \bigcap V_k = V_{mk} \neq \{0\}, m \neq k, \quad \bigcap_{\substack{m,k=1,M \\ m \neq k}} V_{mk} = V^K \neq \{0\}.$$

The set $V^K$, which is a part of all layers of MLN ($N^K$ is the number of elements of $V^K$), will be called the kernel of PO multilayer network, and the MLN itself, which has such property, will be called the kernel multilayer network (Fig. 1). Such structures are generated by the general transport system of the Earth and its separate continents, systems of maintenance of vital activity of the city, postal systems, systems of telephone communication, etc. Obviously, the multilayer structure
$$G^K = (V^K, E^K),$$
where $E^K$ is the set of edges between elements of $V^K$ is multiplex.

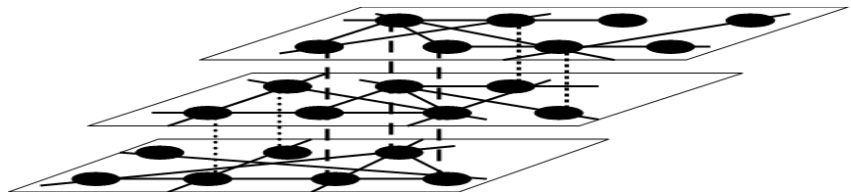

Fig. 1. Example of structure of partially overlapped kernel multilayer network (dashed lines connect the nodes of kernel) without reflecting the direction of connections

   2.2. Partially overlapped multilayer networks in which there is no non-empty intersection of all network layers, ie
$$V_m \bigcap V_k = V_{mk} \neq \{0\}, m \neq k, \quad \bigcap_{\substack{m,k=1,M \\ m \neq k}} V_{mk} = \{0\}.$$

We will call such multilayer network non-kernel. An example of structure of a non-kernel PO MLN is shown in Fig. 2. Such MLN are generated by linguistic multilayer network systems (there is hardly at least one person who speaks all existing languages), social network systems, online services, mobile operators and provider of cable TV, etc [38].

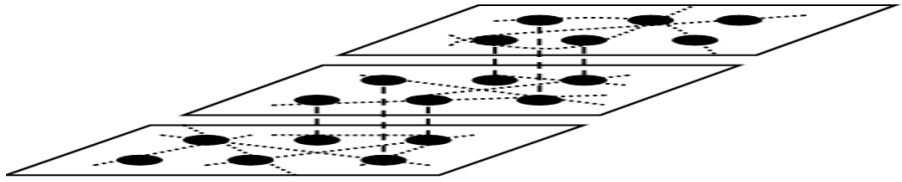

Fig. 2. Example of structure of partially overlapped non-kernel multilayer network network without reflecting the direction of connections

3. Multinetworks, ie multilayer structures in which the sets of nodes of all network layers do not intersect. In this case
$$V_m \cap V_k = \{0\}, \quad m \neq k.$$
Such multilayer structures arise, in particular, during the interaction between producers, carriers, sellers and consumers of certain products and so on.

Denote by $V^P$ the subset of nodes of $V^M$, which belong to more than one layer of MLN, $N^P$ is a number of elements of $V^P$. The relationship between the values $N^M$, $N^K$, and $N^P$ allows us to identify the type of MLN, which is generated by a multilayer network system. Indeed, if condition $N^K = N^P = N^M$ is satisfied, then the structure under study is a multiplex; if $0 < N^K \leq N^P \leq N^M$, then the structure under study is a kernel PO MLN; if $N^K = 0$, $0 < N^P \leq N^M$, then the structure under study is a non-kernel PO MLN. If $N^K = N^P = 0$, then the structure under study is a multinetwork. In general, multiplexes and multinetworks are the limit cases of PO MLN, which occur when the intersection of sets of layers nodes of the latter either forms one network or is an empty set.

According to the method of organization of interlayer connections, we distinguish the following types of MLN structures.
1. Interlayer connections exist only between nodes with the same number from the total set of MLN nodes. From a functional point of view, this means that each MLN node is an element of many systems, in each of which it implements different functions (country in the system of international cooperation), or one function, but in different ways (city as a node of railway, road and air transport systems). From a structural point of view, this means that all nonzero non-diagonal blocks of the adjacency matrix $\mathbf{A}^M$ have a diagonal structure.
2. Each node of a certain layer can have several connections with different nodes of other layers of MLN (control systems in interdependent networks, E-mail and telephone communication systems).

Intersystem interactions and the structures generated by them are dynamic formations. The emergence of new type of interaction between the nodes of networks that are part of it, creates a new layer of MLN. On the contrary, if some kind of interaction disappears, then the corresponding layer disappears. Usually, a new layer and the appropriate type of interaction expands the capabilities of existing connections in multilayer network (fixed and mobile telephone, post and E-mail, etc.). Over time, not only the structure but also the type of MLN may change.

## 2.2. Functional features of multilayer network systems classification

Most real systems and supersystem formations are multipurpose and multifunctional. This is primarily expressed in the multithreading of such formations, ie ensuring the movement of different types of flows [39, 40]. Consider, as an example, a common transport multilayer network system, which includes rail, aviation, road and water (sea and river) transport network layers-systems. Each of these layers provides the movement of flows of two main types: passenger or freight. Therefore, the structure of each of layers-systems of such MLS can be represented in the form of two-dimensional MLN, the edges of each layer of which provide the movement of passenger or freight flows, respectively, and nodes - receiving, processing and sending these flows. Thus, multithreaded network systems should be represented in the form of two-dimensional multilayer systems in which flows of different types to simplify the analysis are divided into different layers-systems. It is obvious that the layers of each of such two-layer transport systems are interacting due to the need to coordinate the movement of carriers of passenger and freight flows. A peculiarity of such MLS is the impossibility of flow transition from one layer to another (conversion of passengers into "freight" or vice versa). The multilayer systems of this type include the above-mentioned systems of international cooperation, ensuring the vital activity of the city, and so on. The impossibility of transforming flows from one type to another in such MLS (ballet troupe into the team of weightlifters and the team of weightlifters into the group of nuclear physicists) means that the process of functioning of separate layers-systems can be modeled and investigated independently.

Obviously, that it is expedient to divide the general transport MLS into two interacting monoflow four-layer network systems, each of which provides the movement of flows of only one type (passenger or freight). A feature of such MLS is the ability to transfer the flow from arbitrary layer to any other. A characteristic peculiarity of monoflow transport MLS is the difference of flow carriers in each layer (trains, vehicles, aircraft, ships). Landline and mobile telephone systems, using the same type of flow, also use different media (wired and wireless). The ability to use different flow carriers not only contributes to the potential increase in their speed or the ability to access network nodes, inaccessible in its separate layers, but also performs the function of duplicating access paths. This can be compared with the duplication of the most important components of complex technological devices or processes that increase the reliability of their work.

Another example of monoflow MLS is social networks and online services for various purposes. Unlike transport and communication, flows in such systems generally have the same carriers, the functioning of which is provided by different systems-operators (Facebook, Twitter, LinkedIn, etc.). In general, during detailing the structure of real multithreaded MLS, it is advisable at first identify the layers that provide the movement of different types of flows, and then each of these monoflow layers to represent in the form of MLS each layer of which provides movement of these flows by a specific carrier or separate operator.

It is obvious that the structural and functional features introduced above allow us to form a fairly clear list of structures of intersystem interactions (or MLS), avoiding some ambiguity and inconsistency of their classification. Hereafter, the subject of our study will be monoflow (MF) partially overlapped multilayer networks, in which interlayer connections are implemented exclusively between nodes with the same numbers in the total set of MLN nodes.

## 3. Structural characteristics of monoflow partially overlapped multilayer network systems

All local (input and output degrees, clustering and branching coefficients, distribution of connection ends, etc.) and global (centralities of different types, availability and expected strength, etc.) characteristics of nodes in the $m$-th layer [41] will be considered its local intralayer characteristics in multilayer network. Let us determine the local characteristics of elements of MF PO MLN that are most important for the analysis of intersystem interactions.

Usually the local or global characteristic of MLN element is considered as the vector of its local or global characteristics in separate system layers [4, 5]. For example, the input and output degrees $\mathbf{d}_i^{in}$ and $\mathbf{d}_i^{out}$ of node $n_i$ in MLN are defined as

$$\mathbf{d}_i^{in} = \{d_i^{m,in}\}_{m=1}^M, \quad \mathbf{d}_i^{out} = \{d_i^{m,out}\}_{m=1}^M$$

where

$$d_i^{m,in} = \sum_{j=1}^{N^M} a_{ji}^{mm}, \quad d_i^{m,out} = \sum_{j=1}^{N^M} a_{ij}^{mm} \qquad (2)$$

determine the input and output degrees of node $n_i$ in the $m$-th layer, respectively, $i = \overline{1, N^M}$. This approach is quite justified in the study of multidimensional or multinetwork systems, since the number of connections of a node in certain layer of corresponding MLN generally does not depend on the number of its connections in other layers. However, when studying the properties of MF PO MLS the local and global structural characteristics can be determined not only for separate layers, but also for a set of interlayer interactions in general.

### 3.1. Aggregate-networks of monoflow partially overlapped multilayer networks

The local characteristic $\varepsilon_{ij}$ of edge $(n_i, n_j)$ in MF PO MLN, which will be called its aggregate-weight, is equal to the number of layers in which such edge is present. Here and below we assume that $i, j = \overline{1, N^M}$. The aggregate-weight $\varepsilon_{ii}$ of node $n_i$ in a partially overlapped multilayer network is equal to the number of layers of which it is a part. A node that belongs to several layers of MLS and through which the flow can move from one layer to another will be called the transition point of multilayer network. For an arbitrary MF PO MLN the adjacency matrix $\mathbf{E} = \{\varepsilon_{ij}\}_{i,j=1}^{N^M}$ ccompletely defines the weighted network, which will be called the aggregate-network of monoflow partially overlapped MLN. The elements of matrix **E** determine the integral structural characteristics of nodes and edges of such multilayer network (Fig. 3).

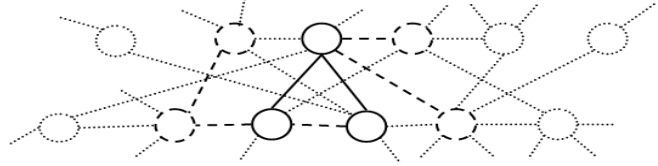

Fig. 3. A fragment of aggregate-network for the shown in fig. 1 partially overlapped kernel three-layer network (____ – for $\varepsilon_{ij}$ =3, _ _ _ – for $\varepsilon_{ij}$ =2, ..... – for $\varepsilon_{ij}$ =1) without reflecting the direction of connections

For multithreaded multidimensional networks, the value of aggregate-weights $\varepsilon_{ij}$, $i \neq j$, of weighted aggregate-network determines the number of interactions of different types between the nodes of such structures. The projection on spatial weighted aggregate-network of multidimensional MLN loses some meaning, because the weight of each of its edge, unlike the multigraph from which such MLN is formed, determines not the type of connection between nodes, but the total number of connections of different types. For monoflow MLN, this disadvantage is absent, because the weight of each edge reflects the number of possible carriers or systems-operators that can provide the movement of corresponding type of flow. For temporal networks, the weighted aggregate network determines the stability of connections between nodes $n_i$ and $n_j$ over time.

Input and output weighted aggregate-degrees of nodes of PO MLN's aggregate-network are determined by the ratios

$$d_i^{\varepsilon,in} = \sum_{j=1}^{N^M} \varepsilon_{ji}, \ d_i^{\varepsilon,out} = \sum_{j=1}^{N^M} \varepsilon_{ij} \qquad (3)$$

respectively. On the basis of weighted aggregate-network, it is possible to build a binary aggregate-network of MLN, which completely determines by the adjacency matrix

$$\mathbf{E}^{bin} = \{\varepsilon_{ij}^{bin}\}_{i,j=1}^{N^M} = \begin{cases} 1, & \text{if } \varepsilon_{ij} \neq 0, \ i \neq j, \\ 0, & \text{if } \varepsilon_{ij} = 0, \ i \neq j, \\ \varepsilon_{ii}, & \text{if } i = j, \end{cases}$$

and defines only the structure of connections in the set of network layers without taking into account the multiplicity of ways to implement network interactions.

Input and output degrees of nodes of PO MLN's binary aggregate-network are determined by the ratios

$$d_i^{bin,in} = \sum_{\substack{j=1 \\ j \neq i}}^{N^M} \varepsilon_{ji}^{bin}, \ d_i^{bin,out} = \sum_{\substack{j=1 \\ j \neq i}}^{N^M} \varepsilon_{ij}^{bin} \qquad (4)$$

It is obvious that

$$d_i^{bin,in} \leq d_i^{\varepsilon,in}, \ d_i^{bin,out} \leq d_i^{\varepsilon,out}$$

and the relations $d_i^{\varepsilon,in}/d_i^{bin,in}$ and $d_i^{\varepsilon,out}/d_i^{bin,out}$ determine the level of duplication of input and output interactions of the node $n_i$, respectively. Indirectly, they also determine the importance of this node in structure of MLS, as duplication is usually subject to elements that implement the most important functions in the system. The clustering coefficient and other local and global structural characteristics of the node in MLN, generated by a partially overlapped multilayer system, are determined from the structure of its binary aggregate-network in the same way as for the nodes of spatial network [42]. It is obvious that the local and global characteristics of elements of PO MLN for multiflow multilayer networks have a significantly other meaning. This is another indication of the need to use the functional features of classification of MLN.

During the study of MF PO MLN there are many practically important problems that can be solved at least in the first approximation, using the concept of its aggregate-network. Such problems include:

1. Determination of the shortest paths through the MLN [43] (change of transport modes can significantly speed up time or reduce the cost of movement of passengers and cargos). The shortest path is built in a binary aggregate-network, the structure of which is described by matrix $\mathbf{E}^{bin}$. When such path is determined, it remains to choose the optimal by certain criteria carriers or system-operators on each edge of this path, the aggregate-weight of which exceeds 1.

2. Search for alternative paths of transit flows through different network layers during the isolation of certain zone in separate network layer (use of subway in large cities during traffic jams) [35]. In this case, the nodes and edges that lie in the isolated zone of separate layer are removed from the structure of aggregate-network.

3. Counteracting the spread of epidemics or computer viruses, which due to multilayer interactions can spread much faster than in one layer [44]. Thus, the parameter $d_i^{bin,in}$ determines the number of nodes adjacent to $n_i$ from which the threat of infection of this node may come; parameter $d_i^{bin,out}$ determines the number of nodes adjacent to $n_i$ in which there may be a threat of infection from this node; the aggregate-weight $\varepsilon_{ii}$ of node $n_i$ determines the number of layers of MLN, in which this node can contribute to the spread of infection.

4. Finding the path from arbitrary node of one layer to arbitrary node of another layer, especially if they lie outside the intersection of sets of nodes of these layers. The ability to move a flow from one network layer to another and back through transition points expands access to nodes that are unreachable in separate network layers and allows communication between unconnected components of such layers (for example, the movement of traffic flows across the oceans or to remote regions of separate countries - the Far East in Russia, the northern regions of Canada, the central regions of Australia), etc.

Using the aggregate-network to solve the above problems means reducing the dimensionality of source model by *M* times. Despite this, all above problems are algorithmically complex and resource-intensive even for aggregate-networks of real MLN and require the use of latest methods of data processing [45, 46].

## 3.2. Local characteristics of interlayer interactions

Blocks $\mathbf{A}^{mm}$ of adjacency matrix $\mathbf{A}^M$ describe the structure of intralayer interactions in the *m*-th layer of MLN. Since the loop connections in MLN layers are excluded, ie the diagonal elements of matrices $\mathbf{A}^{mm}$ are equal to zero, the parameters

$$s_{ij}^{in} = \sum_{m=1}^{M} a_{ji}^{mm}, \quad s_{ij}^{out} = \sum_{m=1}^{M} a_{ij}^{mm} \qquad (5)$$

determine the strength of input and output interactions or the number of ways to implement this interactions between nodes $n_i$ and $n_j$ in multilayer network. Then parameters

$$s_i^{in} = \sum_{j=1}^{N^M} s_{ji}^{in} = d_i^{\varepsilon,in}, \quad s_i^{out} = \sum_{j=1}^{N^M} s_{ij}^{out} = d_i^{\varepsilon,out} \qquad (6)$$

where $d_i^{\varepsilon,in}$ and $d_i^{\varepsilon,out}$ are calculated by means of formulas (3), determine the total or aggregate-strength of the input and output interactions of node $n_i$ in the MLN. Parameters

$$d_i^{in} = \sum_{j=1}^{N^M} \frac{s_{ij}^{in}}{\max(1, s_{ij}^{in})} = d_i^{bin,in}, \quad d_i^{out} = \sum_{j=1}^{N^M} \frac{s_{ij}^{out}}{\max(1, s_{ij}^{out})} = d_i^{bin,out} \qquad (7)$$

determine the total number of input and output connections of node $n_i$ in all layers of MLN or input and output aggregate-degree of this node in the multilayer network, respectively. In other words, the values $d_i^{in}$ and $d_i^{out}$ determine the number of adjacent to $n_i$ nodes in the MLN in general. Obviously that

$$d_i^{in} \geq d_i^{m,in}, \ d_i^{out} \geq d_i^{m,out},$$

and equality is achieved when the node $n_i$ is a part of only one layer of PO MLN. If the input and output aggregate-degrees $d_i^{in}$ and $d_i^{out}$ determine only the number of adjacent nodes of this node in all layers of MLN, the weighted aggregate-degrees take into account the number of such connections in different layers, ie variants of interactions between node $n_i$ and adjacent nodes in multilayer network.

The local characteristics of elements of multilayer network introduced by formulas (5)-(7) show a close connection between the properties of MLN and its aggregate-network. However, in the aggregate-network, the properties of transition points are encapsulated or integrated, determining only the number of layers of which they are part. That is, aggregate-networks can not describe all structural properties of intersystem interactions.

Matrices $\mathbf{A}^{mk}$, $m \neq k$, describe the structure of interlayer interactions in PO MLN. Since we consider monoflow systems in which the interlayer interactions are possible only between nodes with the same numbers in the total set of nodes, matrices $\mathbf{A}^{mk}$ have a diagonal structure and diagonal elements $a_{ii}^{mk}$ of these matrices can be different from 0 if and only if node $n_i$ is part of *m*-th and *k*-th layer of MLN, $m \neq k$.

Parameters

$$\delta_i^{m,in} = \sum_{k=1}^{M} a_{ii}^{km}, \ \delta_i^{m,out} = \sum_{k=1}^{M} a_{ii}^{mk}, \qquad (8)$$

determine the input and output interlayer degrees of node $n_i$ in the *m*-th layer of the MLN. Next, the parameters

$$\delta_i^{in} = \sum_{k=1}^{M} \delta_i^{m,in}, \ \delta_i^{out} = \sum_{k=1}^{M} \delta_i^{m,out}, \qquad (9)$$

determine the total or aggregate-degree of the input and output interlayer interactions of node $n_i$ in the MF PO MLN, respectively.

Global characteristics of a node in such MLN, for example, its centralities of different types in TCN are defined [5], as a vector of node centralities in separate layers of MLN. That is, the importance of node on a certain indicator of centrality changes during the transition from layer to layer. This approach is quite acceptable for multidimensional MLS, in which the transition of flow from layer to layer is generally impossible. However, for monoflow partially overlapped multilayer systems, given the capability of direct or indirect interinteraction between any two nodes of connected structures generated by systems of this type, it is possible to determine the global characteristics of nodes for MLN in general. Thus, the degree centrality of node $n_i$ in monoflow partially overlapped multilayer network is determined by the values of its input $d_i^{in}$ and output $d_i^{out}$ aggregate-degrees calculated by formulas (7). Using a binary aggregate-network, the betweenness centrality of a given node in MLN can be defined as the ratio of the sum of all shortest paths passing through this node to the sum of all shortest paths of binary aggregate-network of MLN. The eigenvalue centrality of MLN nodes can be determined by finding the eigenvalues of adjacency matrix of binary aggregate-network, and so on. This means that the importance of nodes in MF PO MLN will generally be redistributed and may differ significantly from these values in each particular layer.

### 3.3. Structural characteristics of layers of monoflow partially overlapped multilayer networks

The global structural characteristics of multilayer network generated by the MF PO MLS can be considered the number of network layers, as well as dimension, density, diameter, average shortest path length, total and average clustering coefficients of corresponding binary aggregate-network of this MLN. All similar characteristics of the *m*-th layer will be considered its local characteristics in a multilayer network. Let's define some general and most important for an analysis of intersystem interactions the characteristics of layers of MF PO MLN which values it is expedient to consider at construction and research the models of multilayer structures:

1) the specific weight $\theta_m$ of the set of nodes of the *m*-th layer in the total set of nodes of MLN, which is determined by the ratio

$$\theta_m = \frac{N_m}{N^M};$$

it is obvious that if the value $\theta_m \ll 1$, then this layer at least initially can not be included in the model of studied structure;

2) the specific weight $\vartheta_m$ of the set of edges of the *m*-th layer in the total set of edges of MLN, which is determined by the ratio

$$\vartheta_m = \frac{L_m}{L^M};$$

it is obvious that if the value $\vartheta_m \ll 1$, then this layer at least initially can not be included in the model of studied structure;

3) input and output degrees of the *m*-th layer, which are equal to the number of transition points to this layer and from this layer to all other layers of the MLN or the total number of its input and output interlayer interactions, which are determined by the parameters

$$D_m^{in} = \sum_{i=1}^{N_m} \delta_i^{m,in}, \ D_m^{out} = \sum_{i=1}^{N_m} \delta_i^{m,out}$$

for which the values $\delta_i^{m,in}$ and $\delta_i^{m,out}$, $i = \overline{1, N_m}$, are calculated by formulas (8);

4) the specific weight of transition points of the *m*-th layer in the set of all transition points of MLN, which determines the accessibility of interlayer interactions for this layer.

If necessary, we can introduce the characteristics of input and output interlayer interactions between two concrete layers of MLN, etc.

## 4. *k*-cores of aggregate-networks of monoflow partially overlapped multilayer networks

Henceforth, to simplify the presentation, we assume that

$$a_{ij}^{mk} = a_{ji}^{mk}, \ a_{ij}^{mk} = a_{ij}^{km}, \ i,j = \overline{1, N^M}, \ m,k = \overline{1, M},$$

that is, the studied PO MLN are undirected. Then the aggregate-weights of edges of aggregate-network

$$\varepsilon_{ij} = \varepsilon_{ji}, \ \varepsilon_{ij}^{bin} = \varepsilon_{ji}^{bin},$$

and all input and output intra- and interlayer characteristics of MLN elements, defined by formulas (2)-(9), are equal, ie

$$d_i^{\varepsilon,in} = d_i^{\varepsilon,out} = d_i^\varepsilon, \ d_i^{bin,in} = d_i^{bin,out} = d_i^{bin},$$

$$s_{ij}^{in} = s_{ij}^{out} = s_{ij}, \ s_i^{in} = s_i^{out} = s_i,$$

$$d_i^{in} = d_i^{out} = d_i,$$

$$\delta_i^{m,in} = \delta_i^{m,out} = \delta_i^m, \ \delta_i^{in} = \delta_i^{out} = \delta_i.$$

In TCN, one of the ways to determine the structurally most important components of MLN and simplify the models of multilayer networks is the concept of **k**-core [47]

$$\mathbf{k} = \{k_1, k_2, ..., k_M\},$$

as a combination of $k_m$-cores of separate layers of MLN. The values $k_m$ for different layers may differ, which can be explained by the following example. In the transport MLN of Ukraine, the degrees of nodes of railway and road transport networks generally do not exceed 7. At the same time, the degrees of some nodes of aviation (Kyiv, Kharkiv) and water (ports South, Mariupol) transport networks reach many dozens. Therefore, the introduction of notion, for example, **10** = {10, 10, 10, 10}-core of this MLN excludes from its structure much more important for the country railway and road layers. Moreover, **k**-cores with the same values $k_m = k$ can potentially exclude from structure the nodes that are important for organization of intersystem interactions (Chita in Russia, Lunnan in China, Santaren in Brazil and many more). In general, the **k**-core determines the elements that are structurally important more for the MLN layers than for organization of interlayer interactions in it.

It was shown above that the values of many characteristics of MLN elements and its aggregate-network are the same. Therefore, to determine the structurally most important components of MLN, we can use the concept of *k*- cores of its binary and weighted aggregate-networks. Under $k_{ag}^{bin}$-core of binary aggregate-network we understand its largest subnetwork, the degree of nodes $d_i^{bin}$ of which are not less than $k_{ag}^{bin}$. Under $k_{ag}^{\varepsilon}$-core of weighted aggregate-network we understand its largest subnetwork, the degree of nodes $d_i^{\varepsilon}$ of which are not less than $k_{ag}^{\varepsilon}$. From inequality

$$d_i^{bin} \leq d_i^{\varepsilon}$$

follows that the structure of $k_{ag}^{bin}$-core is a component of $k_{ag}^{\varepsilon}$-core in the case $k_{ag}^{\varepsilon} \leq k_{ag}^{bin}$. Indeed, let us the node $n_i$ be part of two layers of MLN, in each of which its degree is equal to 4 and the sets of adjacent nodes in different layers coincide. Then the weighted aggregate degree of this node $d_i^{\varepsilon} = 8$, but its binary aggregate degree $d_i^{bin} = 4$, ie node $n_i$ is a part of $5_{ag}^{\varepsilon}$-core, but not part of $5_{ag}^{bin}$-core. A node $n_j$ may also belong to two layers of the MLN, but in one of them its degree is equal to 2, and in the other - 3, and the sets of adjacent to $n_j$ nodes in different layers do not intersect. Then $d_j^{bin} = d_j^{\varepsilon} = 5$ and this node belong to both of the above 5-cores of aggregate-network. At the same time, if $k_m = k_{ag}^{bin}$ then structure of $k_m$-core is component of $k_{ag}^{bin}$-core.

Adjacency matrices $\mathbf{E}(k_{ag}^{\varepsilon})$ and $\mathbf{E}^{bin}(k_{ag}^{bin})$, which fully describe the structures of weighted $k_{ag}^{\varepsilon}$- and binary $k_{ag}^{bin}$-cores of PO MLN aggregate-network, are obviously obtained from adjacency matrices $\mathbf{E}$ and $\mathbf{E}^{bin}$ by excluding rows and columns for nodes whose values $d_i^{\varepsilon} < k_{ag}^{\varepsilon}$ and $d_i^{bin} < k_{ag}^{bin}$, $i = \overline{1, N^M}$, respectively. Investigation of properties and methods of application the $k_{ag}^{bin}$-cores of binary aggregate-network is carried out similarly [48, 49]. It should also be noted that $k_{ag}^{\varepsilon}$-core of weighted aggregate-network of PO MLN provides much more important information for the study of intersystem interactions. Indeed, if two nodes of aggregate-network have the same degree $d_i^{bin}$, but different values $d_i^{\varepsilon}$, the more important in MLN will be a node with a larger value $d_i^{\varepsilon}$, because the greater this value, the more layers this node belongs to or the more its connections with adjacent nodes in different layers of multilayer network. That is, if the value $d_i^{bin}$ reflects only the number of adjacent to $n_i$ nodes, then $d_i^{\varepsilon}$ takes into account the amount of ways to implement interactions with adjacent nodes.

## 5. Structural *p*-cores of monoflow partially overlapped multilayer networks

To solve the problem of determining the structurally most important components of intersystem interactions in MLS, we introduce the notion of *p*-core of partially overlapped multilayer network

$$\widetilde{G}^p = (\widetilde{V}^p, \widetilde{E}^p),$$

as combination of those subnetworks of separate layers with connections between nodes of them that are part of at least $p$, $2 \leq p \leq M$, layers of MLN (it is obvious that $\widetilde{G}^1$ is identical to the source multilayer network $G^M$ defined by formula (1)). In other words, the *p*-core of MLN is its multilayer subnet, each of the nodes of which has an aggregate-weight $\varepsilon_{ii} \geq p$. Obviously that

$$\widetilde{G}^{p+1} \subset \widetilde{G}^p, \, p = \overline{1, M-1}.$$

If

$$p_{\max} = \max_{i=1, N^M}\{\varepsilon_{ii}\} = M,$$

that is, if the structural *M*-core of multilayer network is nonempty, then the studied MLN is kernel (Fig. 4) and the kernel $\widetilde{G}^M$ has a multiplex structure.

If the condition

$$p_{\max} < M,$$

is met then the studied MLN is non-kernel (Fig. 5).

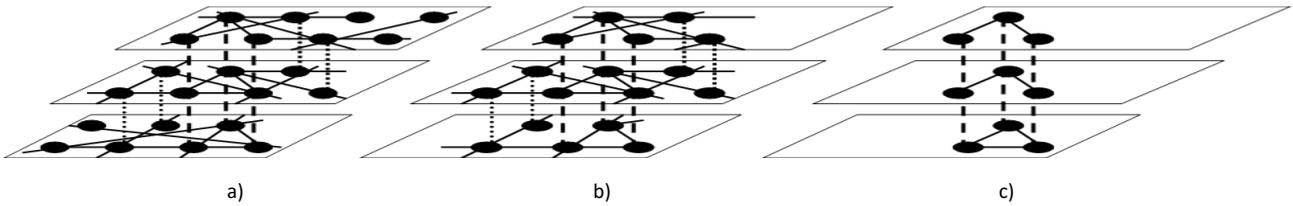

a)                  b)                  c)

Fig. 4. Examples of reflected in fig. 1 source three-layer partially overlapped kernel MLN (a) and its *2*- (b) and *3*-core without reflecting the direction of connections

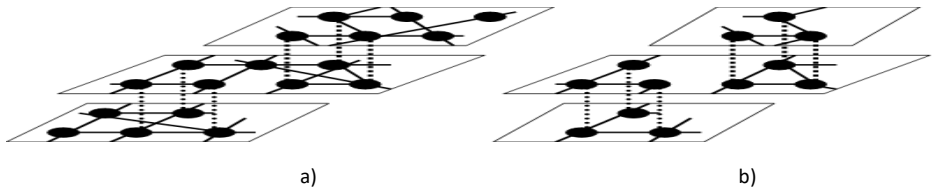

a)                  b)

Fig. 5. Examples of reflected in fig. 2 source three-layer partially overlapped non-kernel MLN (a) and its *2*-core without reflecting the direction of connections

Hereinafter, the $p_{ag}$-core of aggregate-network of MLN will be called such its weighted subnet, each node $n_i$ of which has an aggregate weight $\varepsilon_{ii}$ not less than *p* (Figs. 6 and 7).

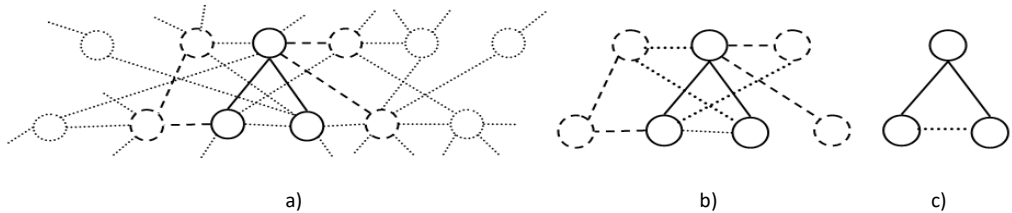

a)                  b)                  c)

Fig. 6. Examples of weighted aggregate-network of reflected in fig. 4 source three-layer partially overlapped kernel MLN (a) and its $2_{ag}$- (b) and $3_{ag}$-core (c, ⎯⎯ – element is part of three layers, - - - - – element is part of two layers, ……. – element is part of one layer) without reflecting the direction of connections

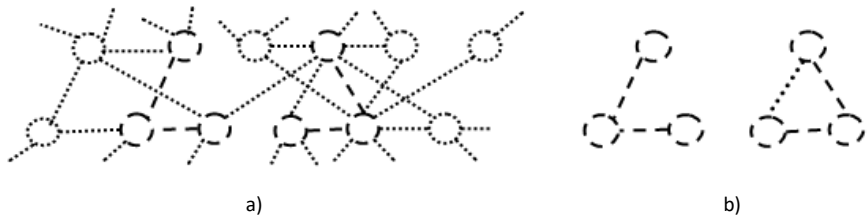

a)                  b)

Fig. 7. Examples of weighted aggregate-network of reflected in fig. 5 source three-layer partially overlapped non-kernel MLN (a) and its $2_{ag}$-core (b, - - - - – element is part of two layers, ……. – element is part of one layer)

Sometimes the question arises as to which networks and in what sequence should be included in the structure of multilayer network. In general, it depends on the purpose of study of intersystem interactions, which are reflected in this structure. Thus, domestic passenger and freight transportation in Ukraine is provided mainly by rail and road. The share of domestic transportation carried out by air and water, as well as nodes of the relevant networks, does not exceed a few percent. This means that when studying the interaction of transport systems (TS) within Ukraine, we can initially limit ourselves to a two-layer network generated by railway and road transport systems. At the same time, the **10**-core of general TS of Ukraine will degenerate into multiplexes, which will contain no more than 10 nodes (compared to more than 1600 nodes of railway TS and 40,000 nodes of road TS). In other countries of the world the situation may be radically different. Thus, the volume of passenger and freight traffic by river transport through the Amazon basin in Brazil, the Nile in Sudan or the Mekong in Laos and Cambodia is commensurate with, and sometimes exceeds, the volume of traffic carried out in these countries by other types of transport networks. However, in the study of interstate passenger traffic, it is impractical to neglect the air TS, and in the analysis of freight traffic - water transport, because their volumes are proportional or exceed the volume of interstate rail and road transport.

Denote by $N^p$ the number of elements of the total set of nodes of the p-core of partially overlapped multilayer network. Let us determine the structural specific weight $\eta^p$ of p-core transition points in this MLN by the ratio

$$\eta^p = \frac{N^p}{N^M}.$$

Obviously that

$$N^{p+1} \leq N^p,$$

and therefore,

$$\eta^{p+1} \leq \eta^p, \quad p = \overline{2(1)M}.$$

The *p*-cores determine the set of transition points that are directly involved in the organization of intersystem interactions at least of *p* layers of MLN. Therefore, the closer the value $\eta^p$ to 1 for all *p = 1,2,…, M,* the stronger the interaction of MLN layers. The specific weight of transition points of the layer determines its ability to participate in intersystem interactions with other layers of the multilayer network. However, this does not mean that only transition points can be involved in intersystem interactions. Other nodes of the layer can participate in such interactions, using transition points as transit. Thus, many Ukrainian agricultural producers, metallurgical, chemical or defense companies export their products through seaports, delivering its in them by rail. Sometimes such nodes are more involved in intersystem interactions than transition points, which act as intermediaries or transiters of flows between layers-systems.

The value $N^M$ only partially reflects the complexity of multilayer network. Much more adequate indicators are the characteristics of dimensional and connectional complexity of MLN. The dimensional complexity of partially overlapped MLN can be estimated through the diagonal elements of matrix **E** by means of parameter

$$\varphi = \sum_{i=1}^{N^M} \varepsilon_{ii},$$

and its connectional complexity – through the elements of matrix **E** using parameter

$$\phi = \sum_{\substack{i,j=1 \\ i \neq j}}^{N^M} \varepsilon_{ij}.$$

To calculate the corresponding indicators of complexity of the *p*-core of MLN, we define the matrix

$$\mathbf{E}_p = \{\varepsilon_{ij}^p\}_{i,j=1}^{N^M}, \quad \varepsilon_{ij}^p = \begin{cases} \varepsilon_{ij}, & \text{if } \varepsilon_{ii} \geq p, \\ 0, & \text{if } \varepsilon_{ii} < p, \end{cases} \quad i,j = \overline{1, N^M}, \quad p = 2,3,...,M.$$

Then the dimensional complexity of *p*-core of partially overlapped MLN we determine by means of parameter

$$\varphi_p = \sum_{i=1}^{N^M} \varepsilon_{ii}^p,$$

and its connectional complexity - through the elements of matrix $\mathbf{E}_p$ using parameter

$$\phi_p = \sum_{\substack{i,j=1 \\ i \neq j}}^{N^M} \varepsilon_{ij}^p, \quad p = 2,3,...,M.$$

Based on this, we can determine the reduction of dimensional complexity of the *p*-core model compared to the PO MLN model in general by the ratio

$$\eta_p = \frac{\varphi_p}{\varphi},$$

and reduction of connectional complexity of the *p*-core model compared to the MF PO MLN model in general by the ratio

$$\mu_p = \frac{\phi_p}{\phi}, \quad p = 2,3,\ldots,M.$$

As in the case of *k*-cores, the use of *p*-cores of partially overlapped multilayer networks has some disadvantages. Thus, the nodes that indirectly participate in intersystem interactions are excluded from its structure. In order to evaluate the adequacy of reduced by means of *p*-core the model of MLS structure we can only on the base of its structural indicators. In addition, *p*-cores determine the criterion for community detection in PO MLS [50, 51]. Indeed, as the value *p* increases, the connections between nodes of *p*-core become denser due to the combination of corresponding subnets that are part of this core, as well as stronger due to the summarizing for a certain period of time the flow volumes that passed through the edges of *p*-core included in its structure. Sometimes such communities are defined even visually (Fig. 6 and 7).

Intersystem interactions can be of different nature, both positive and negative. By positive we mean interactions that improve the state, operation and contribute to the goal of formation and existence of corresponding multilayer system. Examples of positive intersystem interactions are interdisciplinary scientific research, comprehensive implementation of new technologies, expansion of interstate cooperation, successful overcoming of natural and man-made disasters, epidemics of dangerous infectious diseases, and so on. By negative we mean interactions that worsen the functioning of a separate layer or multilayer system as a whole (industrial society has a detrimental effect on the Earth's climate and ecology, the spread of fakes distorts public opinion, etc.). In some multilayer system, depending on the conditions in which it is, intersystem interactions can be both positive and negative. An example is the motor transport system of large city, the separate layers of which are different trucking companies-carriers. If the volume of passenger traffic is greater than or equal to the posibilities of these carriers, the main task of such MLN is to organize effective interaction between them for provision of optimal transportation. Such actions are generally positive. If the volume of passenger flows is less than posibilities of carriers, then there may be competition, which contributes to cheaper and better quality of transportation, ie is a positive factor, or conflictness, which can even lead to armed clashes between separate carriers [52]. It is obvious that *p*-cores of MLN, especially with increasing *p* values, allow us to identify areas that can be called zones of intersection of interests of its various layers-systems (Figs. 4, 5). Depending on the nature of interactions in such zones, they may require either coordination and optimization of joint actions, or measures to avoid potential conflicts.

Note that $k_{ag}^{\varepsilon}$- and $k_{ag}^{bin}$-cores are formed on the basis of values of the sum of nondiagonal elements of the rows or columns of matrices $\mathbf{E}$ and $\mathbf{E}^{bin}$, respectively. At the same time, *p*-cores are formed on the basis of values of the diagonal elements of matrix $\mathbf{E}$. This means that there is no direct connection between the k- and p-cores. However, their properties can be combined as follows. To determine the sets of nodes with the largest values of degree centrality, which at the same time take the greatest part in process of intersystem interactions, we introduce the notion of $p_{ag}(k_{ag}^{bin})$-core as a such subnet of source aggregate-network, nodes of which at first have aggregate-degree $d_i^{bin} \geq k_{ag}^{bin}$ and at second belong to not less than *p* layers of MLN. The adjacency matrix $\mathbf{E}(p_{ag}(k_{ag}^{bin}))$ of corresponding subnet of source MLN is obviously obtained from the adjacency matrix $\mathbf{E}^{bin}(k_{ag}^{bin})$ by excluding rows and columns whose diagonal elements are smaller than $p_{ag}$. Similarly, the notion of $p_{ag}(k_{ag}^{\varepsilon})$-core of the source multilayer network is defined.

Conversely, to determine the sets of nodes that participate most in the process of intersystem interactions and among them have the greatest values of aggregate-degree centrality, we introduce the concept of $k_{ag}^{bin}(p_{ag})$-core as a subnet of source aggregate-network, nodes which at first are the part of at least *p* layers of MLN and at second have the aggregate-degree $d_i^{bin} \geq k_{ag}^{bin}$. The adjacency matrix $\mathbf{E}(k_{ag}^{bin}(p_{ag}))$ of corresponding subnet of the source MLN is obviously obtained from the adjacency matrix $\mathbf{E}_p$ by excluding rows and columns for which the nodes have the aggregate-degree $d_i^{bin} < k_{ag}^{bin}$. Similarly, the notion of $k_{ag}^{\varepsilon}(p_{ag})$-core of the source multilayer network is defined.

It is obvious that the structures of $p_{ag}(k_{ag}^{bin})$- and $k_{ag}^{\varepsilon}(p_{ag})$-cores, as well as the adjacency matrices $\mathbf{E}^{bin}(k_{ag}^{bin})$ and $\mathbf{E}(k_{ag}^{bin}(p_{ag}))$ may differ, because the methods and purposes of construction of each of them are different.

## 6. "Small world" properties of monoflow partially overlapped multilayer networks

Assume that all network layers of multiplex system are "small world" networks [53]. Consider a binary aggregate-network generated by multilayer structure of such monoflow system. Obviously, the density of this aggregate-network is not less than the density of any of network layers of the source multiplex. It is known [54] that the average length of shortest path <*l*> decreases, and the clustering coefficient *C* of connected network increases with increasing its density. Indeed, the value <*l*> is the largest (equal to *N* – 1) and the value *C* is the smallest (equal to 0) for serie of *N* connected nodes. For a complete graph with the same number of nodes, the value <*l*> is the smallest (equal to 1), and the value *C* is the largest (equal to 1). It follows that the average length of shortest path $<l_{agr}>$ in a binary aggregate-network of multiplex, which, as for each layer of "small world" network, includes *N* nodes, is not greater and the clustering coefficient $C_{agr}$ of this aggregate network is not less than for an arbitrary layer of multiplex. That is, if all layers of the multiplex structure are networks of "small world" and $<l_m>$ and $C_m$ are their average lengths of shortest path and clustering coefficients, respectively, then the next inequalities are true

$$<l_m> \leq <l_{agr}>$$

and

$$C_m \leq C_{agr}, \quad m = \overline{1, M} .$$

This means that the feature of "small world" in multiplex structure, each layer of which has this property, is amplified in the sense of reducing the average length of shortest path and increasing the clustering coefficient of aggregate-network. If we consider the sequence of connected *p*-cores of monoflow kernel or non-kernel partially overlapped MLN, each layer of which is a network of "small world", it is obvious that this feature will consistently increase with increasing *p* value in each of the *p*-cores, $2 \leq p \leq M$. This is often facilitated by the fact that the diameter of aggregate-network of *p*-core of the MLN only decrease with increasing of value *p*.

## 7. Vulnerability of intersystem interactions

The study of stability of real systems and intersystem interactions of different types to targeted attacks is one of the most important problems of modern systems analysis. Let us that all layers of MLN are the free-scale networks [54], which are quite widespread and at the same time the most vulnerable to such attacks. The stability of separate system-layers can be determined using the scenarios proposed in [55, 56]. These scenarios are based on the defeat of network nodes with the highest degree or nodes with the highest betweenness centrality. Stability of MLN to targeted attacks is determined primarily by the vulnerability of transition points of the multilayer system, ie its *p*-cores with different values *p*, starting with the largest $p \leq M$. We will build scenarios of attacks on intersystem interactions, using local and global characteristics of the elements of weighted and binary aggregate-networks, which determine the integrated indicators of their importance in MF PO MLN. Note that usually the removal of a particular node from the system structure leads to redistribution of traffic routes that passed through it by the network, and the establishment of new connections between the remaining nodes. This process reflects the reaction of system to changes in operating conditions and means a change of degrees and betweenness centralities of MLN nodes.

Before constructing scenarios of targeted attacks, it is necessary to determine the criteria for their success, ie the desired level of damage of system structure. Such criteria include the division of aggregate-network into unconnected components in which the MLS ceases to exist as a single supersystem formation, the cessation of intersystem interactions between layers, and so on. Let us build the first attack scenario on the basis of aggregate-weights and weighted aggregate-degrees of nodes of the weighted aggregate-network of MLN:

1) make a list of nodes of the aggregate-network of MLN in descending order of their aggregate-weights (values $\varepsilon_{ii}$ of matrix **E**);

2) in each group of nodes with the same values $\varepsilon_{ii}$, we arrange the nodes on the basis of decreasing values of their weighted aggregate degrees $d_i^\varepsilon$;

3) delete the first node from the beginning of created list; if the selected criterion of attack success is fulfilled than algorithm is finished;

4) the removal of particular node usually leads to the establishment of new connections between the nodes that remain in MLN, ie the structure of edges in different layers of MLN and its weighted aggregate-network, and therefore the values $d_i^\varepsilon$ of nodes may change; therefore, if the list of nodes with the current value $\varepsilon_{ii}$ is not exhausted, we pass to point 2 of this scenario;

5) after processing the nodes of group with the current value $\varepsilon_{ii}$ >1 go to step 2 with the value $\varepsilon_{ii}$ -1; if $\varepsilon_{ii}$ =1, then the algorithm is finished.

The second scenario of targeted attack will be built on the use of aggregate-weights and betweenness centrality of nodes of the binary aggregate-network of MLN:

1) make a list of nodes of the aggregate-network of MLN in descending order of their aggregate weights (values $\varepsilon_{ii}$ of matrix **E**);

2) in each group of nodes with the same values $\varepsilon_{ii}$, we arrange the nodes on the basis of decreasing values of their betweenness centralities in binary aggregate-network of MLN;

3) delete the first node from the beginning of created list; if the selected criterion of attack success is fulfilled than algorithm is finished;

4) the removal of particular node usually leads to the establishment of new connections between the nodes that remain in MLN, ie the structure of edges in different layers of MLN and its binary aggregate-network, and therefore the values of betweenness centrality of nodes may change; therefore, if the list of nodes with the current value $\varepsilon_{ii}$ is not exhausted we pass to point 2 of this scenario;

5) after processing the nodes of group with the current value $\varepsilon_{ii}$ >1 go to step 2 with the value $\varepsilon_{ii}$ -1; if $\varepsilon_{ii}$ =1, then the algorithm is finished.

Usually, scenarios that list the characteristics of elements after removal of the next node and are based on the use of betweenness centrality, are more effective for achieving the goal of attack than scenarios that use degree centrality [57, 58].

Along with described above, there is a reverse problem, which is to prevent the spread of epidemics of dangerous infectious diseases, computer viruses, invasion processes, etc. This problem is especially acute during Covid-19 and is the need to block those components of MLN that most "contribute" to the pandemic spreading [59, 60]. The greatest risk of infection in settlements arises during direct contact of people in places of their mass concentration or constant communication. The spread of infection between settlements, regions and countries is due to large volumes of passenger traffic between them. These features are quite well correlated with the elements of MLN with large values of aggregate weights and weighted aggregate-degrees of nodes of its aggregate-network. Hence the ways to prevent the pandemic spreading, which are to block the nodes where crowds are possible, and to block the paths of flows motion to/from the nodes, which have a high level of infection. Obviously, to implement these ways, we can use the proposed above scenarios, in which the criterion for success of anti-epidemiological measures is to minimize access to the nodes with the highest level of infection.

In [56] was shown that the average performance of Internet is halved if only 1% of nodes-domains with the largest degrees fail, and the Internet becomes divided into unconnected components if 4% of such nodes fail. In Ukraine, the number of state-owned banks in the country's banking system does not exceed 0.7%. At the same time, the share of their assets in this system is 55.2%, and the share of deposits of individuals is 61.6% [61]. A successful attack on this group of banks will lead to the largest losses in financial system of the state. This means that for critical destabilization or shutdown of the system operation, it is usually necessary to simultaneously block the operation of a certain group of nodes. Indeed, successive attacks on separate, even the most important nodes, often allow us to distribute their functions among other nodes of the system. This is duly taken into account in the above scenarios. However, to counteract the simultaneous successful attack on a group of the most important elements of MLS, and the main thing to overcome the consequences of such attack, is much more difficult [62]. To defeat intersystem interactions, such group can be a certain set of the most important transition points. The notions of $p_{ag}$ -, $p_{ag}(k_{ag}^\varepsilon)$ - and $k_{ag}^\varepsilon(p_{ag})$ - cores defined in the previous paragraphs allow us to determine the most important for MLS operation groups of nodes, simultaneous targeted attacks on which will certainly cause the most damage or even lead to the cessation of intersystem interactions.

One of scenarios of such defeat, based on the use of notion of $k_{ag}^{\varepsilon}(p_{ag})$-core of aggregate-network of MLN, consists of sequential implementation of the following steps:

1) we accept equal $p_{ag} = p^{\max} \leq M$ and $k_{ag}^{\varepsilon} = \max\limits_{i=1,N^M} d_i^{\varepsilon}$;

2) remove from structure of MLN nodes that belong to $k_{ag}^{\varepsilon}(p_{ag})$-core; if selected criterion for the attack success is met, the algorithm is finished, otherwise go to the next point;

3) reduce the value $k_{ag}^{\varepsilon}$ by 1; if the list of nodes of $k_{ag}^{\varepsilon}(p_{ag})$-core is not exhausted, then go to point 2, otherwise go to the next point;

4) if the value $p_{ag} > 1$, then reduce it by 1 and go to point 2 with a value $k_{ag}^{\varepsilon}$ equal to the maximum weighted aggregate-degree of remaining nodes of aggregate-network; otherwise the algorithm is finished.

To counteract the epidemic spreading, it is also advisable to use scenarios of blocking groups of nodes (settlements) in which the highest incidence is observed. Thus, the practice of preventing the spread of Covid-19 in Ukraine has shown that sequential blocking of the most infected nodes (settlements) gives a worse result than simultaneous blocking groups of nodes (regions of the country) from which the infection can potentially spread [63].

Usually, the spread of epidemics (cholera in Yemen (2016), dengue fever in Sri Lanka (2017), Ebola fever in West Africa (2014) and Congo (2018) and many others) is stopped by the introduction of quarantine, ie the isolation of areas where carriers of infection are found. From a functional point of view, the isolation of certain subnet of the source network means the complete cessation or significant restriction of flows motion from (in, through) it [64]. For many reasons, measures taken to combat the deployment of Covid-19 pandemic have transformed the world into network of isolated zones, movement of flows between which (especially human ones) reduced by dozens due to the cessation or significant restriction of rail, air and road services. Moreover, as a result of introduction of such restrictions, many states have also become networks of isolated communities or separate settlements. The self-isolation of majority of citizens, caused by traffic restrictions, large fines for non-compliance with quarantine conditions and the closure of enterprises or their operation in remote access, has significantly reduced not only external but also internal flows in such isolated zones. With the constant network structure there was a kind of "granulation" of the system, which was divided into a hierarchy of successively isolated subsystems in terms of limiting the interaction between them. Thus, a new type of "granular" networks has emerged, which has not been studied so far. In this case, the losses suffered by the system are not caused by blocking its separate component, but by restricting the movement of flows between all components as a whole. The possibility of transition of network system to the "granular" state, which is characterized by complete or partial isolation of all its components, creates a new kind of stability problems, the feature of which is the defeat not of its separate elements, but of the system in general.

## 8. Conclusions

The study of real complex networks and intersystem interactions of different types allows us to understand many processes that take place in the physical world, nature and human society. The main obstacle that arises in this way is the problem of complexity, caused both by dimension of such systems and the number of heterogeneous interactions in them. One way to solve this problem is to highlight the most important from structural and functional point of view components of system that determine its behavior. To simplify the study of intersystem interactions, the article introduces the notion of aggregate-network, by means of which determines the local and global integral characteristics of elements of multilayer network, and its $k$-core, which highlights the most structurally important components of MLN. To determine the most important for implementation of intersystem interactions components of MLN, the notion of its $p$-core is defined. It is shown how the notions introduced in the paper can be used to solve a number of practically important problems of the theory of complex networks, in particular the stability of intersystem interactions to various vulnerable factors of both artificial and natural origin. As the emergence of new threats, such as Covid-19, is not excluded, the development of scenarios for such defeats and means of timely counteraction to them is an extremely important applied problem, for the theoretical and practical solution of which a lot of effort will be spent.

**Acknowledgments**
Author acknowledges the Director of Pidstryhach Institute for Applied Problems of Mechanics and Mathematics NASU academician Roman Kyshnir and the Head of Laboratory of Modeling and Optimization of Complex Systems DSc Mykhailo Yadzhak for support of this work.

**Data availability statement**
All data that support the findings of this study are included within the article (and any supplementary files).

**ORCID iD**
Olexandr Polishchuk  https://orcid.org/0000-0002-0054-7159